\documentclass[copyright,creativecommons]{eptcs}
\providecommand{\event}{ACL2 2025} 
\usepackage{breakurl}             
\usepackage{underscore}           
\usepackage{color}
\usepackage{fancyvrb}
\usepackage{float}
\floatstyle{boxed}
\restylefloat{figure}

\newcommand{\hrefdoc}[2]{\href{https://www.cs.utexas.edu/users/moore/acl2/manuals/latest/index.html?topic=#1}{\underline{#2}}}

\newcommand{\hrefdoctt}[2]{\hrefdoc{#1}{\texttt{#2}}}

\definecolor{darkcyan}{cmyk}{1.0,0.2,0.2,0.2}

\title{Extended Abstract:\\Mutable Objects with Several Implementations}
\author{Matt Kaufmann
\institute{University of Texas at Austin (retired), Austin, TX, USA}
\email{kaufmann@cs.utexas.edu}
\and
Yahya Sohail
\institute{University of Texas at Austin, Austin, TX, USA}
\email{yahya@yahyasohail.com}
\and
Warren A. Hunt, Jr.
\institute{University of Texas at Austin, Austin, TX, USA}
\email{hunt@cs.utexas.edu}
}
\def\titlerunning{Attach-stobj}
\def\authorrunning{M.~Kaufmann, W.~Hunt, Y.~Sohail}
\begin{document}
\maketitle




This extended abstract outlines an ACL2 feature,
\hrefdoctt{ACL2\_\_\_\_ATTACH-STOBJ}{attach-stobj}\footnote{Underlined
  links are to ACL2 documentation topics.  In particular, the topic
  for \hrefdoctt{ACL2\_\_\_\_ATTACH-STOBJ}{attach-stobj} has details
  not included in this abstract.}, that first appeared in ACL2 Version
8.6 (October, 2024).  Familiarity is assumed here with single-threaded
objects, or {\em \hrefdoc{ACL2\_\_\_\_STOBJ}{stobj}s}~\cite{stobjs}
--- not only ordinary concrete stobjs but also
\hrefdoc{ACL2\_\_\_\_DEFABSSTOBJ}{abstract
  stobjs}~\cite{abstract-stobjs}.

For a worked example that illustrates \texttt{attach-stobj}, see the
directory \texttt{demos/attach-stobj/} in the
\hrefdoc{ACL2\_\_\_\_COMMUNITY-BOOKS}{community
  books}~\cite{community-books}, starting with file
\texttt{README.txt} in that directory.  Performance is addressed in
subdirectory \texttt{mem-test/} of that directory and is discussed
in Section~\ref{performance} below.

\section{Background and Acknowledgments}

The evolving x86 model~\cite{fmcad14} (in community books directory
\texttt{projects/x86isa/}) currently represents its memory using an
abstract stobj that is nested in \texttt{X86}, the abstract stobj it
uses to represent its state.  Linux has been booted on this model and
Linux jobs have been run on it.  But for efficient execution for a
variety of applications, we wanted the x86 model to be flexible by
permitting different memory models to be used with it.  This paper
describes an enhancement to ACL2 that permits such substitution of
memory models without requiring recertification of the book that
defines the \texttt{X86} abstract stobj.  We thank Sol Swords for
helpful design feedback; ForrestHunt, Inc. for supporting the research
reported herein; and the reviewers for helpful feedback on this paper.

\section{Overview}

The idea is to allow an abstract stobj \texttt{ST} to be defined in a
book as an {\em attachable} stobj, using keyword argument {\tt
  :attachable t} as shown below, so that different ACL2 sessions can
specify different ways to execute operations on \texttt{ST} without
the need to recertify the book that defines \texttt{ST}.  In
particular, those overriding executions can be available without
re-proving the theorems that have been proved about \texttt{ST}.

Let's outline how this works.  We start with a book, \texttt{B\_ST},
that contains a
\hrefdoc{ACL2\_\_\_\_DEFABSSTOBJ}{\texttt{defabsstobj}} event
introducing the stobj, \texttt{ST}, followed by some theorems.

\begin{Verbatim}[commandchars=\\\{\}]
(defabsstobj ST
  ...
  :attachable t)    ; allows execution of ST to be modified; see below
{\em{;;; ... some theorems ...}}
\end{Verbatim}

\noindent Then, one or more other books could look as follows, where
\texttt{IMPL} is given the same sequence of \texttt{:logic} functions
in its primitives as those of \texttt{ST}.

\begin{verbatim}
(defabsstobj IMPL ...) ; Might be top-level, but might be from an included book
(attach-stobj ST IMPL) ; IMPL may be attached to a stobj ST, introduced later.
(include-book "B_ST")  ; Define ST using :attachable t, so that execution with
                       ;   ST is performed as specified by IMPL.
\end{verbatim}

\noindent The relevant notions and required order are as follows, as
illustrated above.

\begin{itemize}

\item The {\em implementation} stobj, \texttt{IMPL}, is defined
  before the {\tt attach-stobj} event specifying that it will be {\em
    attached to} the {\em attachable} stobj, \texttt{ST}.

\item The {\tt attach-stobj} event precedes the \texttt{defabsstobj}
  event that defines \texttt{ST}.

\end{itemize}

\noindent The user documentation for
\hrefdoctt{ACL2\_\_\_\_ATTACH-STOBJ}{attach-stobj} provides detailed
requirements for its use.  The key idea is to replace the
\texttt{:foundation} and the \texttt{:exec} fields of the attachable
stobj with those of its implementation.

Suppose that \texttt{IMPL} is to be used only as suggested above ---
that is, it will only serve as an implementation stobj to be attached
to some other stobj (for example, \texttt{ST} above).  Then the
\texttt{defabsstobj} event for \texttt{IMPL} can save space by
specifying \texttt{:non-executable t}.  That option's sole effect is
to prevent the creation of a global \texttt{IMPL} stobj (which however
can be created later, if needed, using
\hrefdoc{ACL2\_\_\_\_ADD-GLOBAL-STOBJ}{add-global-stobj}).  Option
\texttt{:non-executable t} is also useful for a stobj if it will only
be used as a child of a superior stobj or as a
\hrefdoc{ACL2\_\_\_\_LOCAL}{local} stobj.

For the motivating application described in the preceding section,
\texttt{ST} and \texttt{IMPL} represent memories within a superior
\texttt{X86} stobj; see
\hrefdoc{ACL2\_\_\_\_NESTED-STOBJS}{nested-stobjs}.
\texttt{Attach-stobj} works as expected when attaching to the child
stobj (whether concrete or abstract), essentially as follows.

\begin{verbatim}
(defabsstobj IMPL ...) ; implementation memory stobj
(attach-stobj ST IMPL) ; IMPL to be attached later to ST
(defabsstobj ST        ; attachable memory stobj
  ...
  :attachable t)
(defabsstobj X86 ...)  ; includes ST as a child, which executes as IMPL
\end{verbatim}

As the new capability was being designed, it was considered to support
introduction of an attachable stobj without \texttt{:exec} fields for
its primitives.  But that would have required developing a semantics
for such ``incomplete'' abstract stobj definitions as well as
modifying the checks.  In particular, what should be done about the
correspondence function and theorems?  There were also problems
involving signatures of exported functions when omitting
\texttt{:exec} fields.  It thus seemed reasonable to make the usual
requirements for abstract stobjs even when they are attachable, which
pertain even to \texttt{:exec} fields that might not ultimately be
used in execution.  One can view local witnesses in
\hrefdoc{ACL2\_\_\_\_ENCAPSULATE}{encapsulate} events as providing a
sort of precedent.

\section{Performance}
\label{performance}

This section illustrates how \texttt{attach-stobj} can provide
substantial performance benefits without incurring extra proof work.
It summarizes results reported in directory
\texttt{demos/attach-stobj/mem-test/} of the
\hrefdoc{ACL2\_\_\_\_COMMUNITY-BOOKS}{community
  books}~\cite{community-books}, in particular its file
\texttt{README.txt}.  We start with the following table, which gives
runtime and physical memory used in the six runs described below.

\begin{center}
\begin{tabular}{c c c c}
Memory & Benchmark & Time (secs) & Size (bytes) \\
\hline
{\it symmetric}  & low  & \verb| 2.75| & \verb|2000085072| \\
{\it symmetric}  & high & \verb| 2.75| & \verb|2000085072| \\
{\it asymmetric} & low  & \verb| 0.00| & \verb|6663495760| \\
{\it asymmetric} & high & \verb|87.91| & \verb|6666641488| \\
{\it attached}   & low  & \verb| 0.00| & \verb|6899818576| \\
{\it attached}   & high & \verb|89.04| & \verb|6902964304| \\
\end{tabular}
\end{center}

\noindent The rows are based on doing 100,000 writes of byte value 1
to random addresses in a range of $2^{30}$ contiguous addresses.
Those addresses start at address 0 for ``low'' writes and at
$6*2^{30}$ for ``high'' writes.  The writes are done after loading
memory models using the following ACL2 commands.

\begin{itemize}

\item {\em symmetric}: \\{\small
\verb|(include-book "centaur/bigmems/bigmem/bigmem" :dir :system)|}

\item {\em asymmetric}: \\{\small
\verb|(include-book "centaur/bigmems/bigmem-asymmetric/bigmem-asymmetric" :dir :system)|}

\item {\em attached}: \\{\small
\verb|(include-book "centaur/bigmems/bigmem-asymmetric/bigmem-asymmetric" :dir :system)| \\
\verb|(attach-stobj bigmem::mem bigmem-asymmetric::mem)| \\
\verb|(include-book "centaur/bigmems/bigmem/bigmem" :dir :system)|}

\end{itemize}

\noindent In all cases, we see that the {\em symmetric} memory model
performs best of the three when writing to ``high'' memory, while the
other two memory models perform best when writing to ``low'' memory.
It can thus be beneficial to choose different memory models for
different applications.

Naive implementations would simply create the {\em symmetric} and {\em
  asymmetric} models and prove desired theorems about each.  But with
\texttt{attach-stobj}, we need only prove theorems about the {\em
  symmetric} model: with \texttt{attach-stobj} one can {\em attach}
the {\em asymmetric} model to the {\em symmetric} model when one wants
the performance provided by the {\em asymmetric} model.

This use of \texttt{attach-stobj} has benefit beyond avoiding the
duplication of proofs.  Books that include the {\em symmetric} model
can be used with either model, depending on whether or not the {\em
  asymmetric} model is first included and then attached to the {\em
  symmetric} model, as shown in the three commands displayed above for
the {\em attached} usage.  Without the availability of
\texttt{attach-stobj}, one would need to develop two such books: one
that includes the {\em symmetric} model and one that includes the {\em
  asymmetric} model.

Note that the performance penalties are minor for using {\em attached}
instead of {\em asymmetric}.  This is a small price to pay for the
benefits described above.

\section{Implementation Notes}

The basic implementation idea for attachable stobjs is reasonably
straightforward.  The \texttt{attach-stobj} event populates a table,
\texttt{attach-stobj-table}: it associates an attachable stobj name,
which must not yet be defined, with an implementation stobj name,
which must already be defined.  Then when the attachable stobj is
later defined, its corresponding implementation stobj is found by
looking in the table --- recursively, since the value may itself have
an attachment.

\begin{verbatim}
(defun attached-stobj (st wrld top)
; Top is t for a top-level call, nil otherwise.
  (let ((st2 (cdr (assoc-eq st (table-alist 'attach-stobj-table wrld)))))
    (cond (st2 (attached-stobj st2 wrld nil))
          (top nil)
          (t st))))
\end{verbatim}

\noindent The main ACL2 source function for implementing
\texttt{defabsstobj}, which is \texttt{defabsstobj-fn1}, calls itself
one time recursively when \texttt{:attachable t} is supplied,
essentially by replacing the \texttt{:exec} (execution) fields and
\texttt{:foundation} of the attachable stobj with those of the
implementation stobj.  But first it checks that those two abstract
stobjs (attachable and implementation) have the same sequence of
\texttt{:logic} fields.  See the source code for the many details
omitted here, in particular, the definition of the function
\texttt{defabsstobj-fn1} mentioned above and the comment entitled
``Essay on Attachable Stobjs''.

The trickiest part is to install the proper code for execution.  When
including a certified book that defines an abstract stobj, code for
that stobj's primitives is normally provided by the book's compiled
file, based on the \texttt{:exec} fields of the \texttt{defabsstobj}
event.  But when the stobj is attachable and an attachment is
provided, the \texttt{:exec} fields of the implementation stobj need
to be used instead.  This presents a challenge, especially since
abstract stobj primitives are macros.  Consider for example a book
\texttt{B\_ST.lisp} that contains the following events.

\begin{verbatim}
(defabsstobj ST
  ...
  :exports (... (p :logic p$a :exec p$c) ...)
  :attachable t)
(defun f (ST) (declare (xargs :stobjs ST)) ... (p ...) ...)
\end{verbatim}

\noindent Now suppose we attach a stobj \texttt{IMPL} to \texttt{ST}
before including that book.

\begin{verbatim}
(defabsstobj IMPL
  ...
  :exports (... (p{impl} :logic p$a :exec p{impl}$c) ..))
(attach-stobj ST IMPL)
(include-book "B_ST") ; defines ST and (f ST); see above
\end{verbatim}

\noindent A naive implementation would load compiled code from the
book \texttt{B\_ST} when it is included; and since \texttt{p} is a
macro, the compiled code for \texttt{f} would be produced by
macroexpanding calls of \texttt{p} by calling \texttt{p\$c}.  But
instead, those calls of \texttt{p\$c} should instead be calls of the
corresponding \texttt{:exec} field from \texttt{IMPL}, namely,
\texttt{p\{impl\}\$c}.

This problem is addressed using two globals in the logical world,
\texttt{ext-gens} and \texttt{ext-gen-barriers}, that track functions
like \texttt{f} for which compiled code from a book should be ignored,
so that primitives of an attached stobj are invoked using their
attachments.  This will generally cause a function like \texttt{f} to
be compiled when its ACL2 definitional event is encountered during
\hrefdoctt{ACL2\_\_\_\_INCLUDE-BOOK}{include-book}.  Details are
beyond the scope of this extended abstract.  However, that and other
implementation issues are covered in the Essay mentioned above.  The
community books directory \texttt{system/tests/attachable-stobjs/} has
examples that test aspects of attachable stobjs, including some of the
trickier aspects of execution that involve them.

We conclude by noting that execution with attachable stobjs is
efficient, in that \texttt{attach-stobj} introduces no indirection.
The trade-off is that compilation is performed at
\texttt{include-book} time when existing compiled code is avoided, as
discussed above.

\bibliographystyle{eptcs}
\bibliography{attach-stobj}
\end{document}